\documentclass[runningheads]{llncs}
\usepackage[T1]{fontenc}
\usepackage{graphicx}
\usepackage{amsmath}
\usepackage{booktabs}   % \toprule, \midrule, \bottomrule, \cmidrule
\usepackage{multirow}   % merged table headers
\usepackage{algorithm}
\usepackage{algorithmicx}
\usepackage{algpseudocode}
\usepackage{bbding}     % \Envelope -- marks the corresponding author
\usepackage{color}
\usepackage{hyperref}

\makeatletter
\@ifundefined{discintname}{\def\discintname{Disclosure of Interests.}}{}
\@ifundefined{credits}{\newenvironment{credits}{\small}{}}{}
\makeatother
\begin{document}
\title{DiTAR+: Dual Optimization for Robust Autoregressive
Diffusion Speech Synthesis}
%
% Abbreviated title for the running head.
\titlerunning{DiTAR+: Dual Optimization for AR-DiT Speech Generation}
%
% =====================================================================
% Authors, affiliations and e-mail addresses.
%
% RULES (llncsdoc v2.26, Sect. 3.2-3.3):
%   * Affiliations are numbered AUTOMATICALLY from the \and separators
%     inside \institute -- never hand-write $^1$ / $^2$.
%   * \email{...} must sit INSIDE \institute, not after it.
%   * Given name always precedes the family name.
%   * Springer allows at most ONE corresponding author; it is marked
%     here with the envelope symbol (\Envelope).
%   * If several authors share an affiliation, the order of the e-mail
%     addresses must follow the order of those authors.
%
% =====================================================================
% NOTE (llncs 规则)：
%   * \email{} 必须写在 \institute{} 内部。llncs.cls 中 \institute 只把内容
%     存进 \@institute 宏（由 \maketitle 延后排版），而 \email 是
%     \def\email#1{{\tt#1}} —— 会立即在当前位置排版。放到 \institute 外面
%     会先在正文里打出邮箱，再被 \@maketitle 的 \newpage 把标题推到第 2 页。
%   * 只有一个单位时不要写 \inst{1}：\institutename 在单一单位情形下不给
%     单位加编号，作者名上的上标 "1" 会找不到对应项。
\author{Ziyu Zhang \and
Tianlun Zuo \and
Hanzhao Li \and
Haoyu Zhang \and
Lei Xie\textsuperscript{(\Envelope)}}
%
% Given names are shortened to initials in the running head; 'et al.' is
% used because the paper has more than two authors.
\authorrunning{Z. Zhang et al.}
\institute{ASLP@NPU, Northwestern Polytechnical University, Xi'an, China\\
\email{ziyu\_zhang@mail.nwpu.edu.cn, lxie@nwpu.edu.cn}}
\maketitle              % typeset the header of the contribution
\begin{abstract}
Continuous-latent Autoregressive Diffusion Transformer (AR-DiT) models
have demonstrated immense potential in zero-shot speech generation.
However, they still suffer from limited decoding stability when
synthesizing long utterances or complex linguistic structures. This
instability primarily stems from a restricted historical receptive field
and an acoustic inertia dependency within the diffusion decoder, which
causes the model to ignore semantic conditions. To address these
challenges, we propose DiTAR+, a dual-optimization framework. First, we
introduce Dilated Context Sampling to expand the macro-level historical
receptive field without violating physical temporal continuity, thereby
preventing cumulative error propagation. Second, we propose Hierarchical
Acoustic Masking to prevent shallow layers from attending to acoustic
pre-context, explicitly decoupling semantic alignment from acoustic
detail reconstruction. Extensive experiments show that our framework
effectively mitigates pronunciation errors and semantic hallucinations,
enhances generation robustness on challenging sentences, and maintains
exceptionally high speaker similarity throughout the entirety of
long-form utterances. On the linguistically challenging ZH-Hard set,
DiTAR+ reduces the word error rate from 12.478\% to 9.893\%, and on
extended utterances of 25 to 35 seconds it improves speaker similarity
from 0.741 to 0.759 while simultaneously lowering the word error rate
from 2.778\% to 2.173\%, outperforming both discrete-token and pure
flow-matching baselines.

\end{abstract}
\section{Introduction}
Recent advances in speech generation have shown a clear trend from
discrete acoustic token modeling toward continuous representation
modeling~\cite{huang2023make,liu2023audioldm,liu2024audioldm,shen2024naturalspeech}.
Compared with discrete codec tokens, continuous acoustic latents
preserve richer prosodic, timbral, and fine-grained spectral details,
thereby offering a higher upper bound for natural and expressive speech
synthesis~\cite{zhou2025transfusion,ju2024naturalspeech,le2023voicebox}.
Built upon this direction, autoregressive diffusion Transformer
architectures, which combine the sequence modeling ability of causal
language models with the high-fidelity generation capability of
diffusion decoders, have demonstrated strong potential in zero-shot and
long-form speech
generation~\cite{peebles2023scalable,jia2025ditar,zhou2025transfusion,meng2024autoregressive}.
However, despite their impressive quality, continuous-latent AR-DiT
models still suffer from limited generation stability, especially when
synthesizing long utterances or linguistically difficult sentences. This
instability is mainly reflected in speaker similarity degradation, word
repetition, pronunciation disorder, insertion, deletion, and semantic
misalignment.

Existing speech generation methods based on autoregressive modeling can
be broadly categorized into two paradigms: discrete-token-based
autoregressive models and continuous-latent autoregressive diffusion
models. Discrete-token-based methods first convert speech into codec
indices and then employ a language model to predict acoustic
tokens~\cite{wang2023neural,borsos2023audiolm,kharitonov2023speak,borsos2023soundstorm}.
This paradigm benefits from the maturity of text language modeling and
usually achieves robust sequence-level prediction. Nevertheless, the
discretization process inevitably removes part of the acoustic
details~\cite{zeghidour2021soundstream,defossez2023high}, making it
difficult to fully preserve subtle speaker characteristics, prosodic
variations, and expressive nuances. In contrast, continuous-latent
AR-DiT models directly operate on compressed continuous speech
representations and employ a diffusion Transformer decoder to
reconstruct local acoustic
segments~\cite{jia2025ditar,zhou2025transfusion,meng2024autoregressive}.
This design avoids the information bottleneck introduced by vector
quantization and enables more realistic acoustic rendering. Among these
models, DiTAR~\cite{jia2025ditar} is a particularly strong baseline
because it introduces an autoregressive language model to predict
compressed speech patches while using a local diffusion Transformer,
namely LocDiT, to generate high-quality continuous acoustic latents
conditioned on text features and local acoustic history. Such a hybrid
design provides a promising balance between long-range sequence modeling
and local high-fidelity acoustic generation.

Despite these advantages, DiTAR-like AR-DiT models still face two
critical limitations. First, the historical receptive field is
inherently restricted by the local prediction scheme---LocDiT relies
mainly on the immediately preceding patch as acoustic pre-context,
offering only a narrow view of the acoustic trajectory. As utterances
lengthen, remote speaker and prosodic information becomes inaccessible,
causing cumulative error propagation and declining speaker consistency.
Second, LocDiT develops an acoustic inertia dependency during iterative
denoising. Shallow layers over-rely on local acoustic pre-context,
taking shortcuts by copying or continuing previous acoustic patterns,
which weakens the role of semantic conditions from the upstream language
model. When handling complex linguistic structures or long dependencies,
the decoder ignores semantic instructions and produces errors such as
repetitions, omissions, incorrect pronunciation, or loop-like
degeneration. Richer historical context only exacerbates this issue, as
it reinforces acoustic continuation over semantic alignment.

In this paper, we propose DiTAR+, a dual-optimization framework for
robust continuous-latent AR-DiT speech generation. The proposed
framework contains two complementary components: Dilated Context
Sampling (DCS) and Hierarchical Acoustic Masking (HAM). First, to expand
the historical receptive field without violating temporal continuity, we
introduce DCS at the autoregressive front end. Instead of reordering the
speech latent sequence, the proposed strategy preserves the immediately
preceding patch for local boundary smoothness while additionally
sampling remote historical patches with a dilated interval. In this way,
the model obtains a wider macro-level acoustic view at negligible
computational cost, while the absolute physical order of continuous
speech latents remains intact. Second, to mitigate the acoustic inertia
dependency of LocDiT, we introduce HAM at the diffusion back end.
Specifically, the shallow layers of LocDiT are prevented from attending
to acoustic pre-context, forcing the model to first establish semantic
alignment based on linguistic conditions. The deeper layers then recover
access to historical acoustic context for speaker-consistent and smooth
acoustic rendering. This hierarchical design explicitly decouples
semantic alignment from acoustic detail reconstruction, thereby
improving robustness on difficult sentences.

The key contributions of our work are summarized as follows:
\begin{itemize}
\item We propose Dilated Context Sampling, a strategy that expands the
historical receptive field without disrupting temporal continuity,
thereby improving long-form generation stability and speaker
consistency.
\item We design Hierarchical Acoustic Masking to suppress shallow
acoustic shortcut learning, effectively decoupling semantic information
from acoustic rendering and reducing the word error rate.
\item We present DiTAR+, a dual-optimization framework that integrates
DCS and HAM, significantly outperforming baselines in challenging
scenarios by achieving robust zero-shot generation on complex sentences
and maintaining stable speaker similarity across long utterances.
\end{itemize}

% Audio samples can be found at:
% \url{https://ziyuzhang020511.github.io/DiTAR_plus/}.

\section{Method}
\subsection{DiTAR Baseline}
\begin{figure}[t]
\centering
\includegraphics[width=0.90\linewidth]{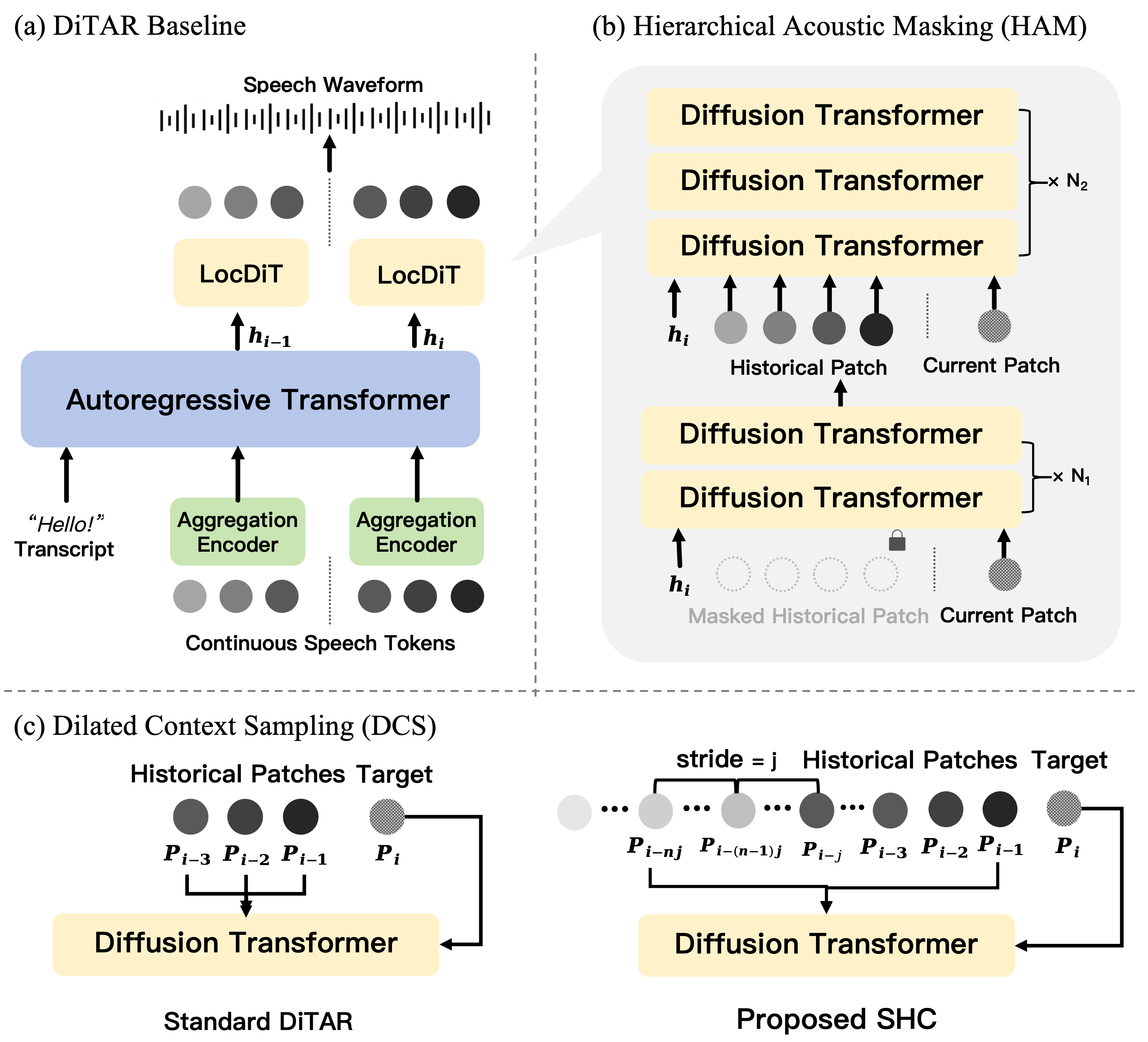}
\caption{The overall architecture of the proposed DiTAR+ framework.
(a) The standard two-stage DiTAR baseline. (b) Hierarchical Acoustic
Masking (HAM), which masks historical acoustic context in shallow layers
to decouple semantic alignment and acoustic rendering. (c) Dilated
Context Sampling (DCS), which expands the macro-receptive field via
dilated sampling while preserving temporal continuity.}
\label{fig:architecture}
\end{figure}

As shown in Fig.~\ref{fig:architecture}(a), the standard DiTAR framework
employs a divide-and-conquer paradigm that seamlessly amalgamates an
autoregressive language model (LM) with a local diffusion decoder
(LocDiT). Initially, the target speech waveform is compressed into a
highly compact continuous latent space via a Variational Autoencoder
(VAE). To facilitate efficient modeling, the continuous tokens are
grouped into non-overlapping patches, each of which is mapped into a
single aggregation embedding. The generation pipeline operates in two
decoupled stages.

\subsubsection{Inter-Patch Semantic Generation.}
At the macroscopic level, a Causal Autoregressive Transformer models the
global sequence dependency. Given the text condition and reference audio
as prefix prompts, the LM processes the aggregated historical latent
patches sequentially. For the $i$-th target patch, the LM
autoregressively predicts a semantic hidden condition vector, denoted as
$h_i$. This vector $h_i$ encapsulates the contextual semantic and
prosodic instructions required for the current patch generation.

\subsubsection{Intra-Patch Acoustic Generation.}
At the microscopic level, a Localized Diffusion Transformer with
bidirectional attention is utilized to reconstruct the fine-grained
continuous speech tokens within each patch. During the iterative
denoising process, LocDiT takes the semantic vector $h_i$, the diffusion
timestep $t$, and the noisy current patch $x_t$ as inputs. Crucially, to
maintain acoustic boundary smoothness, the standard LocDiT explicitly
incorporates the immediate preceding patch (i.e., Patch $i-1$) as a
localized historical context (\texttt{pre\_context}). This
\texttt{pre\_context} is consistently utilized as prefix inputs across
all layers of the diffusion decoder.

\subsection{Dilated Context Sampling}
In the standard DiTAR framework, the local diffusion decoder (LocDiT)
predicts each target patch using only its immediately preceding acoustic
context. Specifically, when generating the $i$-th patch, the model
typically conditions on the $(i-1)$-th patch. While this strictly local
prediction paradigm promotes smooth transitions between adjacent speech
segments, it inherently limits the effective receptive field of the
decoder.

For long-form speech synthesis, such restricted historical context leads
to progressive error accumulation. As generation proceeds, the model
gradually loses access to distant acoustic information, making it
increasingly difficult to maintain speaker characteristics and long-term
acoustic consistency. This issue often manifests as speaker drift, where
speaker similarity deteriorates with increasing utterance length.

A straightforward solution is to enlarge the receptive field through
patch reordering strategies such as
Skip-Patchify~\cite{li2024autoregressive}. However, such approaches
disrupt the original temporal structure of speech representations and
may break local acoustic continuity by separating neighboring patches
that are naturally adjacent in time. To address this limitation, we
propose \textbf{Dilated Context Sampling (DCS)}, as shown in
Fig.~\ref{fig:architecture}(c), a simple yet effective mechanism that
expands the receptive field while preserving the temporal ordering of
acoustic features.

Specifically, when predicting the $i$-th target patch $P_i$, we always
retain a small window of immediately preceding patches (e.g.,
$[P_{i-c}, \dots, P_{i-1}]$) as local context to ensure smooth boundary
transitions. In addition, we employ a dilated sampling strategy to
sparsely gather distant historical patches from earlier positions in the
sequence with a fixed stride $j$. The local continuous context and the
dilated distant context are then concatenated and jointly provided to
LocDiT as conditioning information:
\begin{equation}
C_i = [P_{i-nj}, \ldots, P_{i-2j}, P_{i-j}, P_{i-c}, \ldots, P_{i-2},
P_{i-1}],
\end{equation}
where $C_i$ denotes the overall historical context used for predicting
the $i$-th patch, $c$ is the local context window size (e.g., $c=3$),
$j$ represents the dilation stride (where $j > c$), and $n$ determines
the maximum range of the macro-receptive field.

To explicitly illustrate the data flow of the proposed sampling
strategy, we outline the exact execution steps in Phase 1 of
Algorithm~\ref{alg:locdit_forward}. During the initial phase of each
autoregressive decoding step, the model constructs the context
dynamically rather than relying on a static prefix. As shown in the
algorithm, it sequentially extracts a dense local context window
($C_{\text{local}}$) to maintain boundary smoothness, followed by a
sparsely sampled distant context ($C_{\text{dilated}}$). These two
distinct components are then concatenated to assemble a unified
historical acoustic context ($X_{\text{hist}}$). This programmatic
operation ensures that the macro-receptive field is substantially
expanded while the absolute physical temporal timeline remains strictly
intact prior to being fed into the LocDiT network.

By incorporating sparsely sampled historical patches, DCS enables LocDiT
to simultaneously access both short-range continuity cues and long-range
acoustic information. Unlike patch reordering methods, DCS significantly
enlarges the effective receptive field without altering the original
temporal structure of speech. Consequently, it mitigates long-range
error accumulation, improves speaker consistency in extended speech
generation, and substantially reduces speaker drift while maintaining
high-quality acoustic continuity.

\begin{algorithm}[t]
\caption{Forward pass of LocDiT with DCS and HAM}
\label{alg:locdit_forward}
\textbf{Input:} Current patch index $i$, generated historical patches
$P$, semantic condition $h_i$, current noisy patch $X_{\text{curr}}$. \\
\textbf{Hyperparameters:} Local window size $c$, dilation stride $j$,
max dilated patches $n$, total Transformer layers $L$, masked shallow
layers $N_1$.
\begin{algorithmic}[1]
\Statex \Comment{\textbf{Phase 1: Dilated Context Sampling (DCS)}}
\State $C_{\text{local}} \gets [P_{i-c}, \dots, P_{i-2}, P_{i-1}]$ \Comment{Extract local continuous context}
\State $C_{\text{dilated}} \gets [P_{i-nj}, \dots, P_{i-2j}, P_{i-j}]$ \Comment{Extract sparse distant context}
\State $X_{\text{hist}} \gets \text{Concatenate}(C_{\text{dilated}}, C_{\text{local}})$ \Comment{Assemble context in temporal order}
\Statex
\Statex \Comment{\textbf{Phase 2: Hierarchical Acoustic Masking (HAM)}}
\State $H^{(0)} \gets \text{Concatenate}(h_i, X_{\text{hist}}, X_{\text{curr}})$ \Comment{Construct initial input sequence}
\For{$l = 1$ \textbf{to} $L$}
    \State $Q, K, V \gets \text{LinearProjections}(H^{(l-1)})$
    \State Initialize attention mask $M^{(l)} \gets 0$ for all token pairs
    \If{$l \le N_1$}
        \For{\textbf{each} query $q_u \in X_{\text{curr}}$ \textbf{and} key $k_v \in X_{\text{hist}}$}
            \State $M^{(l)}_{u,v} \gets -\infty$ \Comment{Hide historical acoustic tokens}
        \EndFor
    \EndIf
    \State $\text{Attn\_Out} \gets \text{Softmax}\left(\frac{QK^\top}{\sqrt{d}} + M^{(l)}\right)V$ \Comment{Masked self-attention}
    \State $H^{(l)} \gets \text{FeedForward}(\text{Attn\_Out}) + H^{(l-1)}$ \Comment{Update hidden states}
\EndFor
\Statex
\State \Return Extracted representations corresponding to $X_{\text{curr}}$ from $H^{(L)}$
\end{algorithmic}
\end{algorithm}

\subsection{Hierarchical Acoustic Masking}
In the standard DiTAR architecture, historical token patches are
concatenated as a prefix and directly provided to LocDiT. While this
design improves acoustic continuity, it also introduces a strong
dependency on historical acoustic contexts. Specifically, when the
diffusion model has unrestricted access to the entire pre-context in
early layers, it tends to exploit a shortcut by extrapolating from
previously generated acoustic patterns rather than relying on semantic
guidance from the LM. As a result, semantic conditioning is weakened,
leading to unstable decoding behaviors such as repetitive generation,
looping artifacts, insertion errors, pronunciation mistakes, and a
substantial increase in word error rate, particularly for long-form or
semantically challenging utterances.

To address this issue, we propose a \textbf{Hierarchical Acoustic
Masking (HAM)} strategy within LocDiT. The key idea is to progressively
expose semantic and acoustic conditions across network depth. Let
$X_{\text{curr}}$ denote the current noisy token sequence,
$X_{\text{hist}}$ denote the historical acoustic context (i.e., the
pre-context), and $h_i$ denote the semantic representation produced by
the LM. The masked self-attention operation in the $l$-th Transformer
block is formulated as
\begin{equation}
\text{Attention}(Q,K,V)
=
\text{Softmax}
\left(
\frac{QK^\top}{\sqrt d}
+
M^{(l)}
\right)V,
\label{eq:ham_attention}
\end{equation}
where $M^{(l)}$ is a layer-dependent attention mask. For attention
interactions between query tokens $q_u \in X_{\text{curr}}$ and key
tokens $k_v \in X_{\text{hist}}$, we define
\begin{equation}
M_{u,v}^{(l)}
=
\begin{cases}
-\infty,
&
q_u \in X_{\text{curr}},
\; k_v \in X_{\text{hist}},
\; l \le N_1,
\\
0,
&
\text{otherwise},
\end{cases}
\label{eq:ham_mask}
\end{equation}
where $N_1$ denotes the number of masked shallow Transformer blocks.
Consequently, historical acoustic tokens are completely invisible during
the early stages of denoising, forcing LocDiT to establish semantic
alignment based solely on the LM-provided conditioning $h_i$. In the
subsequent $N_2$ deeper layers, the mask is removed, allowing the model
to incorporate historical acoustic information for fine-grained acoustic
refinement.

To provide a concrete implementation perspective, the layer-by-layer
execution of this dynamic masking mechanism is formally detailed in
Phase 2 of Algorithm~\ref{alg:locdit_forward}. Prior to the Transformer
forward pass, the model seamlessly concatenates the semantic condition
($h_i$), the assembled historical context ($X_{\text{hist}}$), and the
current noisy patch ($X_{\text{curr}}$) into a single input sequence. As
this sequence propagates through the network depth, a conditional loop
dynamically enforces the masking rule: for any layer index $l \le N_1$,
the attention weights corresponding to the historical acoustic keys are
aggressively overridden with $-\infty$. This algorithmic constraint
forcibly cuts off the local acoustic shortcut in shallow layers,
dynamically lifting the restriction only in deeper layers ($l > N_1$) to
finalize the fine-grained acoustic reconstruction.

This progressive conditioning mechanism effectively decouples semantic
alignment from acoustic rendering. By preventing shallow layers from
overfitting to local acoustic continuity, the model is encouraged to
first construct a semantically faithful representation before leveraging
historical acoustic cues for prosody, timbre, and continuity modeling.
Empirically, the proposed masking strategy substantially improves
decoding robustness in challenging long-form synthesis scenarios,
mitigating failure modes such as repetitive loops, insertion and
deletion errors, and pronunciation inconsistencies, while maintaining
high-fidelity speech generation.

\section{Experiments}
\subsection{Experimental Setup}
\subsubsection{Training Data and Hardware.}
All models are trained from scratch on the large-scale multilingual
speech dataset, Emilia~\cite{he2024emilia}. The training process is
conducted on a computing cluster comprising 4 nodes, each equipped with
8 H20 GPUs, totaling 32 H20 GPUs. During training, the batch size is set
to 8 per GPU.

\subsubsection{Inference Configurations.}
The inference process employs an Euler ODE sampler with 10 steps. We
apply Classifier-Free Guidance (CFG) with a scale of $\alpha=2.0$, a
flow-matching scale of $1.0$, and sway-sampling to balance generation
diversity and stability. The maximum prompt length is truncated to 75.
During synthesis, the prompt audio is utilized as the speaker reference
to generate the target text utterance by utterance.

\subsubsection{Evaluation Metrics.}
To comprehensively assess the performance of the generated speech, we
employ both objective and subjective metrics. For objective evaluation,
we measure the Word Error Rate (WER) to reflect linguistic
intelligibility and Speaker Similarity (SIM) to evaluate voice cloning
accuracy, strictly adhering to the standard
\texttt{seed-tts-eval}\footnote{\url{https://github.com/BytedanceSpeech/seed-tts-eval}}
protocol. For subjective evaluation, we conduct multi-dimensional
assessments including Naturalness MOS (N-MOS), Quality MOS (Q-MOS), and
Similarity MOS
(S-MOS)~\cite{itut1996p800,liu2024seed,wang2023neural}. We invited 10
professional evaluators to rate 50 randomly selected utterances per
system on a standard 5-point scale. Additionally, we conduct Comparative
MOS (C-MOS) tests, utilizing our proposed DiTAR+ as the benchmark anchor
(scored at 0.0) to explicitly gauge human preference against other
comparative systems.

\subsubsection{Evaluation Dataset.}
We construct a comprehensive evaluation benchmark covering both standard
and extreme zero-shot synthesis scenarios. First, we adopt the standard
English and Chinese zero-shot test sets from \texttt{seed-tts-eval},
denoted as \textbf{Seed-EN} and \textbf{Seed-ZH}, respectively. To
rigorously evaluate model robustness against complex linguistic
structures, we utilize 400 challenging utterances from the official
\texttt{test-zh-hard} set, denoted as \textbf{ZH-Hard}. Furthermore, to
explicitly assess long-term speaker consistency and the mitigation of
temporal speaker drift, we curate \textbf{ZH-Long}, a specialized subset
comprising 400 extended sentences with expected acoustic durations
ranging from 25 to 35 seconds.

\subsection{Compared Methods}
To comprehensively evaluate the superiority of our proposed DiTAR+
framework, we benchmark it against both our direct baseline and
representative state-of-the-art (SOTA) models covering diverse speech
generation paradigms:
\begin{itemize}
\item \textbf{DiTAR~\cite{jia2025ditar}:} The standard continuous-latent
autoregressive diffusion transformer. We include this to directly
demonstrate the effectiveness of our proposed DCS and HAM modules in
mitigating error accumulation.
\item \textbf{CosyVoice 2~\cite{du2024cosyvoice}:} A leading
discrete-token based autoregressive model, representing the paradigm of
large-scale LM with discrete audio codecs.
\item \textbf{MaskGCT~\cite{wang2025maskgct}:} A state-of-the-art
non-autoregressive (NAR) model utilizing masked generative modeling over
discrete tokens, offering strong bidirectional context understanding.
\item \textbf{F5-TTS~\cite{chen2025f5} and E2 TTS~\cite{eskimez2024e2}:}
Recent highly successful pure continuous flow-matching (diffusion)
models that generate speech without an autoregressive linguistic
planner.
\end{itemize}

\subsection{Objective Evaluation}
The objective evaluation results comparing our proposed DiTAR+ with the
baseline and SOTA models are presented in
Table~\ref{tab:main_objective}. We analyze the performance across
standard evaluation sets and extreme stress-test scenarios.

\subsubsection{Performance on Standard Scenarios.}
On the standard Seed-EN and Seed-ZH subsets, DiTAR+ demonstrates
exceptional fundamental generation capabilities. It achieves the lowest
WER among all evaluated systems, reaching 1.672\% on Seed-EN and 0.985\%
on Seed-ZH. In terms of speaker similarity (SIM), DiTAR+ yields highly
competitive scores (0.735 and 0.762 respectively), performing on par
with or exceeding strong baselines like MaskGCT and the original DiTAR,
confirming its ability to produce highly intelligible and
speaker-consistent speech in normal contexts.

% =====================================================================
% Table 1: Main objective results.
% Table caption is placed ABOVE the table (Springer requirement).
% The table is typeset in \footnotesize instead of being scaled with
% \resizebox, so that the effective font size stays legible and
% consistent with the rest of the paper.
% =====================================================================
\begin{table}[t]
\centering
\caption{Objective evaluation results compared with the baseline and
state-of-the-art models on the SeedTTS test set. WER is reported in \%.
The best result in each column is in bold and the second best is
underlined}
\label{tab:main_objective}
\footnotesize
\setlength{\tabcolsep}{3pt}
\begin{tabular}{lcccccccc}
\toprule
\multirow{2}{*}{\textbf{System}} & \multicolumn{2}{c}{\textbf{Seed-EN}} & \multicolumn{2}{c}{\textbf{Seed-ZH}} & \multicolumn{2}{c}{\textbf{ZH-Hard}} & \multicolumn{2}{c}{\textbf{ZH-Long}} \\
\cmidrule(lr){2-3} \cmidrule(lr){4-5} \cmidrule(lr){6-7} \cmidrule(lr){8-9}
 & \textbf{WER}$\downarrow$ & \textbf{SIM}$\uparrow$ & \textbf{WER}$\downarrow$ & \textbf{SIM}$\uparrow$ & \textbf{WER}$\downarrow$ & \textbf{SIM}$\uparrow$ & \textbf{WER}$\downarrow$ & \textbf{SIM}$\uparrow$ \\
\midrule
Ground Truth & 2.060 & 0.730 & 1.254 & 0.750 & - & - & - & - \\
\midrule
CosyVoice 2 & 2.570 & 0.652 & 1.450 & 0.749 & 13.415 & 0.678 & \underline{2.776} & \underline{0.748} \\
MaskGCT & 2.623 & 0.717 & 2.273 & \textbf{0.774} & 14.085 & \underline{0.688} & 2.954 & 0.740 \\
F5-TTS & 1.982 & 0.670 & 1.560 & 0.760 & \underline{11.209} & 0.669 & 2.894 & 0.729 \\
DiTAR & \underline{1.774} & \textbf{0.737} & \underline{1.287} & 0.753 & 12.478 & 0.672 & 2.778 & 0.741 \\
\midrule
\textbf{DiTAR+} & \textbf{1.672} & \underline{0.735} & \textbf{0.985} & \underline{0.762} & \textbf{9.893} & \textbf{0.695} & \textbf{2.173} & \textbf{0.759} \\
\bottomrule
\end{tabular}
\end{table}

\subsubsection{Robustness on Extreme Scenarios.}
The structural superiority of the proposed dual-optimization framework
becomes prominently evident in the extreme stress-test subsets. On the
linguistically complex \textbf{ZH-Hard} set, the standard DiTAR baseline
suffers from a severe performance drop, yielding a high WER of 12.478\%.
By explicitly decoupling semantic alignment from acoustic rendering via
HAM, DiTAR+ significantly suppresses acoustic hallucination and reduces
the WER to 9.893\%, outperforming all competing models by a large
margin. Furthermore, on the extended-length \textbf{ZH-Long} set, DiTAR+
achieves both the highest SIM (0.759) and the lowest WER (2.173\%). This
validates that the DCS sampling mechanism successfully expands the
macro-receptive field, effectively preventing speaker drift and
cumulative decoding errors during long-form generation without
compromising acoustic continuity.

\subsection{Subjective Evaluation}
We conducted multi-dimensional subjective evaluations with 10 evaluators
on a standard 5-point scale, using DiTAR+ as the C-MOS anchor. The
detailed results are presented in Table~\ref{tab:main_subjective}.

% =====================================================================
% Table 2: Main subjective results.
% =====================================================================
\begin{table}[t]
\centering
\caption{Subjective evaluation results (MOS) on the SeedTTS test set. We
compare DiTAR+ with the DiTAR baseline and two leading flow-matching
systems. The best result in each column is in bold and the second best is
underlined}
\label{tab:main_subjective}
\begin{tabular}{lcccc}
\toprule
\textbf{System} & \textbf{N-MOS$\uparrow$} & \textbf{Q-MOS$\uparrow$} & \textbf{S-MOS$\uparrow$} & \textbf{C-MOS$\uparrow$} \\
\midrule
Ground Truth & 3.88 & 3.60 & 3.57 & +0.22 \\
\midrule
E2 TTS & 3.25 & 3.42 & 3.17 & -0.28 \\
F5-TTS & 3.37 & 3.59 & 3.32 & -0.02 \\
DiTAR & 3.70 & \textbf{3.88} & 3.52 & -0.06 \\
\midrule
\textbf{DiTAR+} & \textbf{3.71} & \underline{3.82} & \textbf{3.58} & \textbf{0.00} \\
\bottomrule
\end{tabular}
\end{table}

\subsubsection{Overall Performance.}
DiTAR+ demonstrates exceptional subjective quality, achieving the
highest Naturalness MOS and Similarity MOS among all evaluated models.
Notably, compared to recent highly successful pure continuous
flow-matching models (e.g., E2 TTS and F5-TTS), our dual-optimization
framework exhibits a clear superiority in capturing and maintaining
speaker characteristics, as reflected in the S-MOS scores. Furthermore,
all competing models yield negative C-MOS scores, underscoring a strong
human preference for the speech synthesized by DiTAR+. Interestingly,
DiTAR+ performs on par with or even marginally exceeds the Ground Truth
in S-MOS. This can be attributed to the model's capacity to generate a
clean, idealized, and stable timbre representation, effectively
filtering out the natural intra-speaker variances (e.g.,
micro-environmental noises) present in real human recordings.

\subsubsection{Analysis of the Q-MOS Trade-off.}
We do note that the standard DiTAR baseline yields a slightly higher
Q-MOS compared to DiTAR+. This represents a deliberate and necessary
architectural trade-off. The standard baseline heavily overfits to the
immediate local acoustic history, which creates locally smooth acoustic
textures (resulting in high Q-MOS) but ultimately causes catastrophic
semantic hallucinations and speaker drift in extended generation. By
integrating DCS and HAM, DiTAR+ explicitly breaks this pathological
local inertia. It trades a negligible loss in micro-level acoustic
smoothness for a substantial leap in macro-level prosodic naturalness,
semantic robustness, and long-term speaker stability.

\subsection{Ablation Study}
\begin{figure}[t]
\centering
\includegraphics[width=0.93\linewidth]{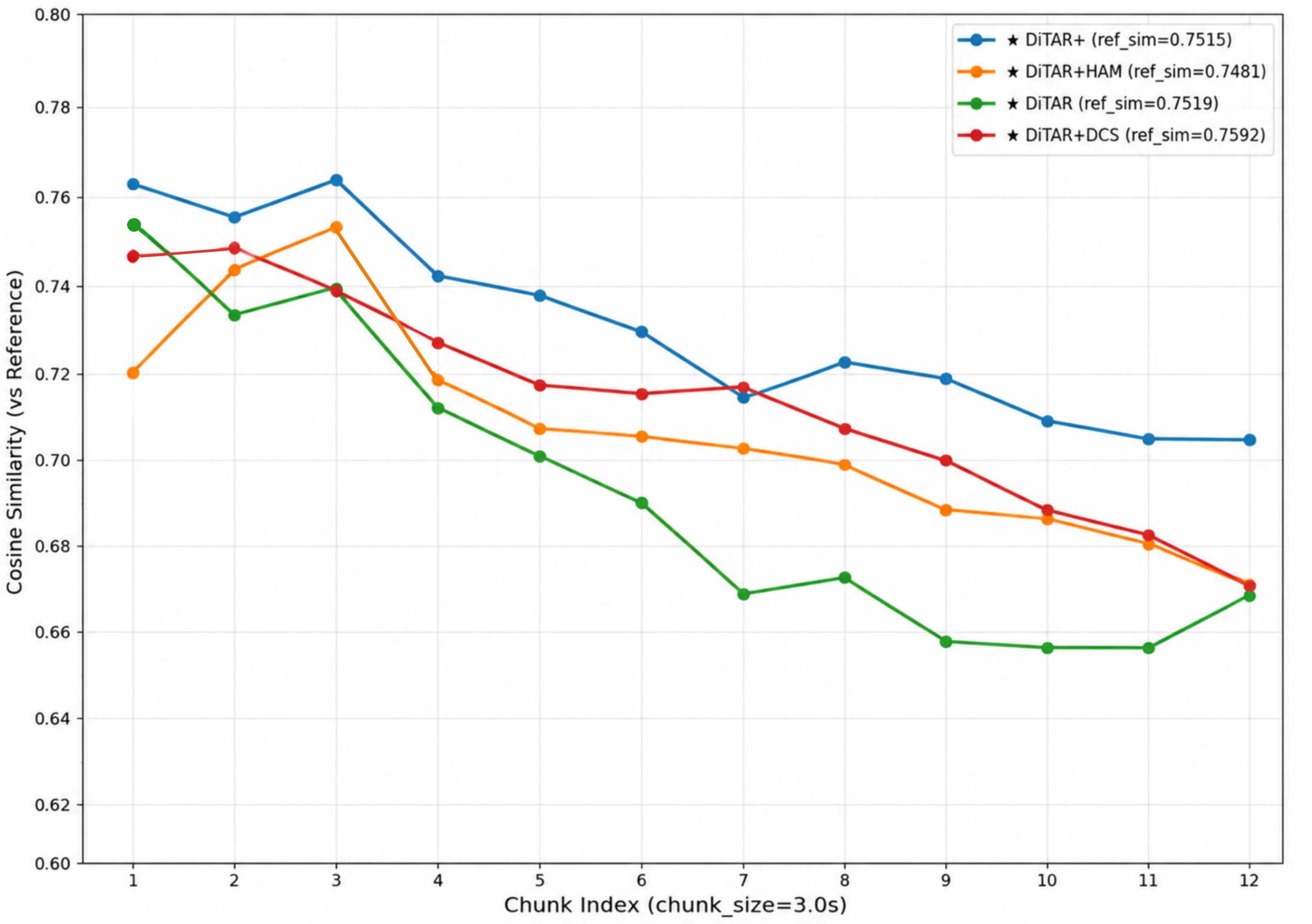}
\caption{Chunk-level speaker similarity over continuous generation steps
(3.0\,s per chunk). The standard baseline experiences severe speaker
drift in the long tail, whereas DCS effectively stabilizes the temporal
coherence.}
\label{fig:ablation_sim}
\end{figure}

To evaluate the individual contributions of our proposed mechanisms, we
conducted an ablation study on the challenging ZH-Hard and ZH-Long sets,
with overall results detailed in Table~\ref{tab:ablation}. Furthermore,
to granularly analyze the temporal dynamics of speaker drift, we
visualize the chunk-level speaker similarity over continuous generation
steps in Fig.~\ref{fig:ablation_sim}.

% =====================================================================
% Table 3: Ablation study.
% =====================================================================
\begin{table}[t]
\centering
\caption{Ablation study of the proposed DCS and HAM modules on the
challenging ZH-Hard and ZH-Long sets. The best result in each column is
in bold and the second best is underlined}
\label{tab:ablation}
\begin{tabular}{lcccc}
\toprule
\multirow{2}{*}{\textbf{System}} & \multicolumn{2}{c}{\textbf{ZH-Hard}} & \multicolumn{2}{c}{\textbf{ZH-Long}} \\
\cmidrule(lr){2-3} \cmidrule(lr){4-5}
 & \textbf{WER(\%)$\downarrow$} & \textbf{SIM$\uparrow$} & \textbf{WER(\%)$\downarrow$} & \textbf{SIM$\uparrow$} \\
\midrule
DiTAR & 12.478 & 0.672 & 2.778 & 0.741 \\
DiTAR + DCS & 10.671 & 0.681 & \textbf{2.165} & \underline{0.757} \\
DiTAR + HAM & \textbf{9.889} & \textbf{0.700} & 3.056 & 0.749 \\
\midrule
DiTAR + Skip & 16.231 & 0.612 & 5.105 & 0.685 \\
\midrule
\textbf{DiTAR+} & \underline{9.893} & \underline{0.695} & \underline{2.173} & \textbf{0.759} \\
\bottomrule
\end{tabular}
\end{table}

\subsubsection{Effectiveness of DCS.}
Integrating the Dilated Context Sampling (\textit{DiTAR + DCS})
primarily enhances long-form generation stability. Compared to the
standard baseline, it significantly reduces the WER on the ZH-Long set
to 2.165\% and boosts overall speaker similarity to 0.757. As clearly
depicted in Fig.~\ref{fig:ablation_sim}, the standard baseline suffers
from severe speaker drift, degrading sharply after the fifth to sixth
chunk (approximately 15 to 18 seconds) and bottoming out around 0.655 at
the tail. In contrast, introducing DCS substantially slows this
degradation, sustaining a markedly higher SIM than the baseline
throughout the extreme tail of long utterances. This visually
demonstrates that DCS successfully expands the macro-receptive field,
effectively mitigating cumulative errors over extended durations.

\subsubsection{Effectiveness of HAM.}
Conversely, applying Hierarchical Acoustic Masking (\textit{DiTAR +
HAM}) yields substantial improvements on linguistically complex
sentences. It achieves the lowest WER (9.889\%) and the highest SIM
(0.700) on the ZH-Hard set, validating that explicitly suppressing
shallow-layer acoustic inertia forces the model to prioritize semantic
alignment. Notably, as shown in Fig.~\ref{fig:ablation_sim}, applying
HAM alone yields a downward SIM trajectory that closely tracks the
baseline in extended generation, converging to a similarly low SIM at
the tail. This aligns with our architectural design: HAM is strictly
formulated to decouple acoustic inertia for semantic accuracy, rather
than extending the historical acoustic context window.

\subsubsection{Failure of Naive Reordering.}
As a negative control, we evaluated a naive non-linear feature
reordering strategy (\textit{DiTAR + Skip}). As expected, this approach
results in severe performance degradation across all metrics, with the
WER soaring to 16.231\% on ZH-Hard and 5.105\% on ZH-Long. This
catastrophic drop provides strong empirical evidence that disrupting the
absolute temporal ordering of continuous latents destroys the learned
temporal prior, rendering the expanded context useless.

\subsubsection{The Full Framework.}
Finally, our complete framework (\textit{DiTAR+}) seamlessly integrates
both modules. By combining the complementary strengths of DCS and HAM,
the full model achieves near-optimal, highly balanced performance across
both extreme scenarios. As visualized in Fig.~\ref{fig:ablation_sim},
the full \textit{DiTAR+} achieves the most stable and highest speaker
similarity trajectory across the entire sequence, suggesting that the
semantic regularization from HAM is largely orthogonal to the temporal
coherence established by DCS.

\begin{figure}[t]
\centering
\includegraphics[width=0.93\linewidth]{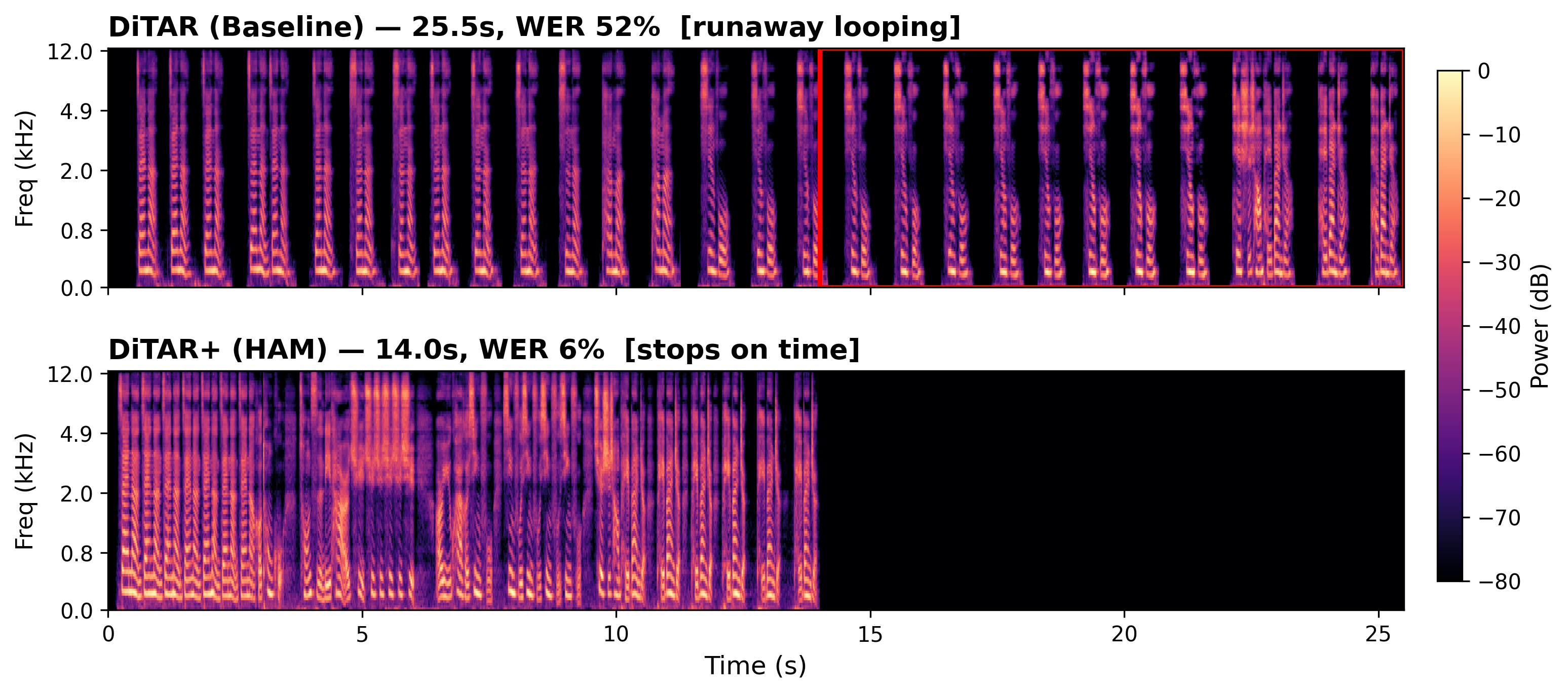}
\caption{Mel-spectrogram comparison between the standard DiTAR baseline
and the proposed DiTAR+ on a challenging utterance. \textbf{Top:} the
baseline model falls into a runaway looping state (highlighted by the
box), failing to terminate generation properly. \textbf{Bottom:} DiTAR+
with the HAM mechanism explicitly suppresses these acoustic shortcuts,
yielding accurate semantic alignment.}
\label{fig:spectrogram_comparison}
\end{figure}

\subsection{Spectrogram Analysis}
To demonstrate the effectiveness of our proposed method in mitigating
semantic hallucinations and decoding instability, we visualize the
Mel-spectrograms of a challenging utterance generated by both the
baseline and our model, as shown in
Fig.~\ref{fig:spectrogram_comparison}.

As depicted in the top panel, the standard DiTAR baseline suffers from
severe ``runaway looping'' during the autoregressive generation process.
Instead of following the semantic guidance from the language model, the
diffusion decoder falls into a state of acoustic inertia, repeatedly
copying previously generated acoustic patterns. This catastrophic
failure results in an abnormally extended audio duration of 25.5 seconds
and an exceptionally high WER of 52\%, with the repetitive artifacts
clearly visible inside the highlighted bounding box.

In contrast, the bottom panel demonstrates the generation result of
DiTAR+ equipped with the HAM mechanism. By aggressively masking
historical acoustic context in shallow layers, HAM forces the model to
strictly align with the semantic condition. As a result, DiTAR+
completely suppresses the looping artifacts, reconstructs clear and
non-repetitive spectral details, and successfully terminates the
generation on time at 14.0 seconds, drastically reducing the WER to
6\%. This qualitative evidence suggests that decoupling semantic
alignment from acoustic rendering is crucial for robust speech
generation.

\section{Conclusion}
In this paper, we proposed DiTAR+, a parameter-free dual-optimization
framework designed to address the decoding instability of
continuous-latent Autoregressive Diffusion Transformer (AR-DiT) models
in zero-shot speech generation. To overcome the fundamental
architectural bottlenecks of restricted receptive fields and acoustic
inertia dependency, we introduced two complementary mechanisms: Dilated
Context Sampling (DCS) and Hierarchical Acoustic Masking (HAM). DCS
effectively expands the macro-level receptive field while strictly
preserving the absolute physical timeline, thereby preventing cumulative
speaker drift in long-form generation. Concurrently, HAM explicitly
decouples semantic alignment from acoustic detail reconstruction by
progressively exposing historical context, significantly mitigating
semantic hallucinations in linguistically complex sentences. Extensive
evaluations demonstrate that by combining the orthogonal strengths of
DCS and HAM, DiTAR+ substantially outperforms state-of-the-art models,
suggesting that temporal phase coherence and semantic-acoustic
decoupling are essential design principles for robust
continuous-latent speech synthesis.

% =====================================================================
% The "credits" environment (llncs v2.23+) switches to 9 pt and turns
% \subsubsection into the small run-in heading Springer asks for.
% \discintname expands to "Disclosure of Interests."
%
% Springer 规范（Instructions for Authors, CS Proceedings）原文：
%   "Disclosure of Interests. It is now necessary to declare any
%    competing interests or to specifically state that the authors have
%    no competing interests."
% 该声明为【强制项】，已启用。
% =====================================================================
\begin{credits}
\subsubsection{\discintname}
The authors have no competing interests to declare that are relevant to
the content of this article.
\end{credits}

%
% ---- Bibliography ----
%
% BibTeX users should specify bibliography style 'splncs04'.
% References will then be sorted and formatted in the correct style.
% Remember to ship the generated .bbl file together with the sources.
%
\bibliographystyle{splncs04}
\bibliography{mybibliography}

\end{document}